\documentclass[a4paper,12pt]{article}
\pdfoutput=1 %

\usepackage[T1]{fontenc}
\usepackage[latin9]{inputenc}
\usepackage[a4paper]{geometry}
\usepackage{verbatim}
\usepackage{amsmath}
\usepackage{stmaryrd}
\usepackage{hyperref}

\usepackage{amsmath}
\usepackage{amssymb}

\usepackage{amsmath} 
\usepackage{url} 
\usepackage{epsfig} 
\usepackage{cite} 
\usepackage{psfrag} 

\usepackage{color}

\newcommand{\nn}{\nonumber}

 \newcounter{RSQ}

\usepackage{slashed}

\begin{document} 
\begin{flushright}
MITP/21-047\\

\end{flushright}

\vskip3cm
\begin{center}
{\Large \bf Refactorisation in subleading $\bar B \to X_s \gamma$}
\end{center}

 \vspace{1.5cm}
\begin{center}
{\sc Tobias Hurth$^{a}$ and Robert~Szafron,$^{b}$} 
\\[6mm]
{\it ${}^a$PRISMA+ Cluster of Excellence and Institute of Physics (THEP),\\
Johannes Gutenberg University, D-55099 Mainz, Germany}\\[6mm]
{\it ${}^b$Department of Physics, Brookhaven National Laboratory, Upton, N.Y., 11973, U.S.A.}
\\[2.3cm]

\end{center}

\begin{abstract}
We establish refactorisation conditions between the subleading ${O}_8$-${O}_8$ contributions to the inclusive $\bar B \to X_s \gamma$ decay suffering from endpoint divergences and prove a factorisation theorem for these contributions to all orders in the strong coupling constant. 
This allows for higher-order calculations of the resolved contributions and consistent summation of large logarithms, consequently reducing the recently found large-scale dependence in these contributions. 
We implement the concept of refactorisation in a heavy flavour application of SCET, which includes nonperturbative functions as additional subtlety not present in collider applications. 
\end{abstract}

\newpage

\tableofcontents

\newpage
\section{Introduction}
There has been a general belief that soft-collinear factorisation at subleading power in $\Lambda_{\rm QCD}/m_b$ expansion is well established for inclusive $B-$decay modes 
such as $\bar B \to X_s \gamma$, $\bar B \to X_s \ell \ell$, or $\bar B \to X_u \ell \bar \nu$~\cite{Beneke:2015wfa} -- in contrast to exclusive $B$ decays where factorisation theorems do not exist at the subleading power in general. 
There are two types of subleading contributions to the inclusive $\bar B \to X_s \gamma$ decay, direct and resolved ones. In the latter, the photon does not directly couple to an effective electroweak vertex, but they contain subprocesses in which the photons couple to light partons instead. These subleading corrections are nonlocal in the endpoint region, and they stay nonlocal even in the region where the local heavy mass expansion is applicable. In this sense, they represent an irreducible uncertainty of this decay mode. Analogous subleading contributions exist in the inclusive $\bar B \to X_s \ell \ell$ decay but not in the inclusive $\bar B \to X_u \ell \bar \nu$ decay because, in this case, the leptons can couple to light partons via the $W$ vector boson only. 

The first systematic analysis of resolved contributions to the inclusive $\bar B \to X_s \gamma$ decay~\cite{Misiak:2015xwa,Hurth:2010tk} was worked out in 
Refs.~\cite{Benzke:2010js,Benzke:2010tq}, the corresponding $1/m_b$ contributions to the inclusive $\bar B \to X_s \ell \ell$ decay were discussed in Refs.~\cite{Hurth:2017xzf,Benzke:2017woq}, using soft collinear effective theory (SCET). Recently, the uncertainty due to the resolved contribution was reduced with the help of a new hadronic input~\cite{Gunawardana:2019gep,Benzke:2020htm}. But these resolved contributions still represent the largest uncertainty in the inclusive $\bar B \to X_s \gamma$ decay. Moreover, a large scale dependence and also a large charm mass dependence were identified in the lowest order result of the resolved contribution, which calls for a systematic calculation of $\alpha_s$ corrections and renormalisation group (RG) summation~\cite{Benzke:2020htm}. A mandatory prerequisite for this task is an all-order in the strong coupling constant $\alpha_s$ factorisation formula for the 
subleading power corrections. 

The factorisation of resolved contributions introduces a new ingredient, namely an anti-hardcollinear jet 
function~\cite{Benzke:2010js}, typically referred to as a radiative or amplitude-level jet function in collider 
and flavour applications~\cite{Lunghi:2002ju,Beneke:2003pa,Bosch:2003fc,Beneke:2018gvs,Beneke:2019oqx,Moult:2019mog,Laenen:2020nrt,Liu:2020ydl,Liu:2021mac}.
They are not represented by cut propagators as the usual jet functions but as full propagator functions (both dressed by Wilson lines). But as already noticed in Ref.~\cite{Benzke:2010js}, the specific resolved ${O}_8 - {O}_8$ contribution does not factorise
because the convolution integral is UV divergent. The authors of Ref.~\cite{Benzke:2010js} emphasised that there is an essential difference between divergent convolution integrals in power-suppressed contributions of exclusive $B$ decays and the divergent convolution integrals in the present case, while the former were of IR origin, the latter divergence of UV nature. However, a solution at the lowest order was established by considering the sum of direct and resolved ${O}_8 - {O}_8$ contributions, which was shown to be scale and scheme dependent by using a hard cut-off in the resolved contribution. But the failure of factorisation does not allow for a consistent resummation of large logarithms. 
 
In this paper, we identify the divergences in the resolved {\it and} in the direct contributions as endpoint divergences by showing that also the divergence in the direct contribution can be traced back to a divergent convolution integral. 
Recently, new techniques were presented in specific collider applications of SCET~\cite{Beneke:2019kgv,Beneke:2019mua,Wang:2019mym,Liu:2019oav,Liu:2020wbn,Beneke:2020ibj, Beneke:2022obx,Liu:2022ajh,Cornella:2022ubo}. The so-called refactorisation conditions or endpoint factorisation\cite{Liu:2019oav,Liu:2020wbn,Beneke:2022obx,Liu:2022ajh,Cornella:2022ubo} allow for an operator-level reshuffling of terms within the factorisation formula so that all endpoint divergences cancel out. In this work, we now implement this idea  in a flavour application of SCET$_{\rm I}$, which includes nonperturbative soft functions not present in collider applications -- often referred to as subleading shape functions~\cite{Beneke:2004in,Bosch:2004cb}.

As a first step, we derive the matching of the hard function for the two operators involved in the ${O}_8 - {O}_8$ 
subleading contributions. 
In the second step, we establish the bare factorisation theorem for the direct and the resolved contribution at an operational level. Then we derive the refactorisation conditions to  all orders, leading us finally to the renormalised factorisation theorem.
We present all steps for the inclusive $\bar B \to X_s \gamma$, but all the details can also be taken over for the corresponding $\bar B \to X_s \ell \ell$ case. 

\vspace{1cm} 

\section{General setup}

The starting point for all calculations concerning the $\bar B \to X_s \gamma$ decay is the weak effective Lagrangian defined at a scale $\mu_b$ parametrically equal to the $b$-quark mass $\mu_b \sim m_b$. The weak effective Lagrangian is obtained from the SM Lagrangian after integrating out the heavy particles like the heavy gauge bosons and the top quark. We use the convention of Ref.~\cite{Beneke:2001ev}. 
Assuming Standard Model CKM unitarity, with $\lambda_q=V_{qb} V_{qs}^*$ and $\lambda_u + \lambda_c + \lambda_t = 0$, the effective Hamiltonian may be written as
\begin{equation}\label{eq:WeakHamiltonian}
  {\cal H}_\text{eff} = \frac{G_F}{\sqrt{2}} \sum_{q=u,c} \lambda_q\,
  \bigg( C_1\,{O}_1^q + C_2\,{O}_2^q+ C_{7\gamma}\,{O}_{7\gamma} + C_{8g}\,{O}_{8g} + \sum_{i=3,...,6} C_i\,{O}_i 
  \bigg) \,.
\end{equation}
Here we concentrate on the ${O}_8$ operator:
\begin{equation}
 {O}_{8g} = -\frac{g_s}{8\pi^2}\,m_b\, \bar s\sigma_{\mu\nu}(1+\gamma_5) G^{\mu\nu} b\,.
\end{equation}  
Our sign convention is that $iD_\mu=i\partial_\mu+g_s\,t^a A_\mu^a+e\,q_f A_\mu$, where $t^a$ are the $SU(3)$ colour generators, and $Q_f$ is the electric charge of the fermion in units of $e$. 
We consider the CP-averaged $\bar B \to X_s \gamma$ photon energy spectrum in the endpoint region where $M_b - 2 E_\gamma = O(\Lambda_{\rm QCD})$. Soft-collinear-effective theory (SCET) offers the appropriate framework for this multi-scale problem. The kinematics of the decay is given as follows: the initial meson carries momentum $p_B$, and it decays into a photon with momentum $q$ and a jet whose total momentum is $P_X$. From $p_B - q = P_X$ in the $B$ meson rest frame, we have $2 M_B E_\gamma = M_B^2 - M_X^2$. Thus, the jet invariant mass $M_{X}$ is much smaller than the photon energy $E_\gamma$ and jet energy $E_X$. We set $P_{X\perp} =0$ and choose reference vectors $n^2=\overline{n}^2=0$, $v^2=1$, such that $n + \bar n = 2v$ and $n \overline{n}=2$. Choosing 
\begin{align}
q^\mu =E_{\gamma}\bar n^\mu \quad {\rm and}  \quad p^\mu_B = M_B v^\mu\,,
\end{align}
we find $M_B = \bar n P_X $ and 
$M_B = n P_X + 2 E_\gamma$ or equivalently 
\begin{equation}
P_{X}^\mu = (M_B - 2E_{\gamma})n^\mu  + M_B \bar n^\mu \,.
\end{equation}
Thus, there is only one independent kinematical variable in the $\bar B \to X_s \gamma$ decay. One may choose the photon energy $E_\gamma$ or $n P_X = M_B - 2 E_\gamma$. 

Three dynamical scales describe the endpoint region, a hard scale of $O(M_B)$, an intermediate
(anti-)hardcollinear scale of $O({\sqrt{ M_B \Lambda_{\rm QCD} }})$, and a soft scale of $O(\Lambda_{\rm QCD})$. 
The expansion parameter in our present analysis is defined as $\lambda^2= \Lambda_{\rm QCD} / M_B$\footnote{Alternative convention often used in the literature is to define $\lambda= \Lambda_{\rm QCD} / M_B$. }.
The photon can be treated as anti-hardcollinear. The hadronic final state factorises into a hardcollinear jet and soft wide-angle radiation.  
Since the soft modes have parametrically smaller virtuality than the hardcollinear modes, the problem at hand is described by the SCET$_{\rm I}$ setup~\cite{Bauer:2000yr,Bauer:2001yt,Beneke:2002ph,Beneke:2002ni}. Using a shorthand notation $a \sim (n a, \bar n a, a_\perp)$ to indicate the scaling of the momentum components in powers of $\lambda$, we have: hard momentum scales like $p_\text{hard} \sim (1, 1, 1)m_b$, a hardcollinear one $p_\text{hc} \sim ( \lambda^2, 1 ,\lambda)m_b$, an anti-hardcollinear region $p_{\overline{\text{hc}}} \sim ( 1, \lambda^2 , \lambda) m_b$ and a soft 
momentum $p_\text{soft} \sim (\lambda^2, \lambda^2, \lambda^2) m_b$.

The first step in the derivation of a factorisation theorem is hard matching. We have to match the electroweak operator onto SCET. We will see that the direct contribution is represented by a next-to-leading power (NLP) $B$-type current in SCET, i.e. power-suppressed current composed of two collinear building blocks.  The resolved contribution is represented by a time-ordered product of a leading-power (LP) $A$-type current with a subleading  $\mathcal{L}_{\xi q}^{\left(1\right)}$ Lagrangian (see Ref.~\cite{Beneke:2017ztn,Beneke:2017mmf,Beneke:2018rbh,Beneke:2019kgv} for a precise definition of the different types of currents). 

In the second step, we integrate out the hardcollinear fields, which lead to the appearance of jet functions. The latter are, technically speaking, matching coefficients of SCET on the pure soft effective field theory.  Kinematics forbids the emission of anti-hardcollinear partons. Thus anti-hardcollinear fields have to be integrated out at the \emph{amplitude} level.
The physics in the anti-hardcollinear direction is similar to the threshold Drell-Yan expansion, where kinematical constraints forbid hardcollinear emissions into the final state. For more details on SCET NLP factorisation and resummation in the  threshold Drell-Yan, see Refs.~\cite{Beneke:2018gvs,Beneke:2019mua,Beneke:2019oqx,Jaskiewicz:2019phy}. In the hardcollinear sector, the formalism resembles, for example, the thrust factorisation, where NLP jet functions are defined at the cross-section level~\cite{Kolodrubetz:2016uim,Feige:2017zci,Moult:2018jjd,Moult:2019mog,Moult:2019uhz}.

\subsection{Hard matching } 

The electroweak operator ${O}_8$ matches onto two possible SCET operators. 
First, we consider the $A$-type current, which enters the resolved contribution. Within the present section, the SCET operators are always given before the soft decoupling transformation~\cite{Bauer:2001yt} is performed. The $A$-type SCET operator is given by 
\begin{equation}\label{eq:A0L}
\mathcal{O}_{8g}^{A0}\left(0\right)=\overline{\chi}_{\overline{hc}}\left(0\right)\frac{\slashed n}{2}\gamma_{\mu\perp}\mathcal{A^{\mu}}_{hc\perp}\left(0\right) \left(1+\gamma_{5}\right) h\left(0\right)\,,
\end{equation}
where $h$ is the heavy quark field, and the SCET building blocks are hardcollinear gauge-invariant due to the introduction of hardcollinear Wilson lines $W$. The fermionic building block is $\chi_{\overline{hc}} =W^\dagger_{\overline{hc}} \xi$ and gluon field is 
\begin{equation}
\mathcal{A^{\mu}}_{hc\perp}=W_{hc}^{\dagger}\left[D_{hc\perp}^{\mu}W_{hc}\right]=\mathcal{A}_{hc\perp}^{a\mu}t^{a}.
\end{equation}
{Note that the colour and Dirac structure for the $A$-type operator is uniquely fixed.}
For the matching, we can use the partonic QCD amplitude $b\left(p_b\right)\to s\left(p_s\right) g\left(r\right)$, where the momenta $p_b$ is hard, $p_s$ anti-hardcollinear and $r$ hardcollinear. 
The matching condition is given by
\begin{equation}
\frac{G_{F}}{\sqrt{2}}\lambda_{t}\,C_{8g}\, \left\langle Q_{8g} \right\rangle = C^{A0}\left(m_{b}\right) \, 
\left\langle {\cal O}_{8g}^{A0} \right\rangle \,.
\end{equation}
The brackets $\left\langle \,\, \right\rangle$ indicate that the matrix element of the operators is considered.
At leading order in $\alpha_s$ we find 
\begin{align*}
C_{LO}^{A0}\left(m_{b}\right) =\frac{m_{b}^{2}}{4\pi^{2}}\frac{G_{F}}{\sqrt{2}}\lambda_{q}C_{8g}\,. 
\end{align*}

The $B$-type current which enters the direct contribution is the following SCET operator:\footnote{Note that this operator is equal to $\mathcal{J}_{2}\left(\tau\right)$
in Ref.~\cite{Beneke:2005gs} (eq. 16), with 
$\mathcal{J}_{2}\left(\tau\right)=2m_{b}\mathcal{O}_{8g}^{B1}\left(\overline{\tau}\right)$.} 
\begin{equation}
\mathcal{O}_{8g}^{B1}\left(u\right)= \int\frac{dt}{2\pi}e^{-ium_{b}t}\overline{\chi}_{hc}\left(t\bar n\right)\gamma_{\nu\perp}\,  Q_s\,\mathcal{B^{\nu}}_{\overline{hc}\perp}\left(0\right)\gamma_{\mu\perp} \, \mathcal{A^{\mu}}_{hc\perp}\left(0\right)\left(1+\gamma_{5}\right) h\left(0\right),\label{eq:OB1}
\end{equation}
with $Q_s$ as the electric charge of the strange quark in units of $e$ and the electromagnetic gauge-invariant transverse photon field 
\begin{equation}
\mathcal{B^{\nu}}_{\overline{hc}\perp} = e\left(A_{\perp}^{\nu}-\frac{\partial_{\perp}^{\nu}}{n\partial} nA\right).\,
\end{equation}
We note that beyond the LO, a second operator with an alternative Dirac structure $\gamma_\mu \gamma_\nu$ appears. These operators do not mix under renormalisation~\cite{Beneke:2005gs}. We checked by explicit computation at the one-loop order that the matching coefficient for the operator with $\gamma_\mu \gamma_\nu$ structure does not develop any endpoint divergence.

We use the partonic QCD amplitude $b\left(p_b\right)\to g\left(r\right)s\left(p_s\right)\gamma\left(q\right)$
to fix the matching coefficient. Here the momenta are $p_b$ hard, $r$ and $p_s$ hardcollinear and $q$ 
anti-hardcollinear. On the QCD side, a time-ordered product of the operator $O_8$ and of the QED current,  
\begin{equation}
\mathcal{L}_{QED,q}\left(x\right)= e_{q}\, A_{\mu}\left(x\right)\, \overline{q}\left(x\right)\gamma^{\mu}q\left(x\right)\,,
\end{equation}
is needed. 
The matching condition at the leading order in QED is given by 
\begin{equation}
\frac{G_{F}}{\sqrt{2}}\lambda_{t}C_{8g}\, i \int d^{4}x\,  \left\langle   T\left[O_{8g}\left(0\right),
\mathcal{L}_{{\small QED},b}\left(x\right)\,
 +\mathcal{L}_{QED,s}\left(x\right)\right] \, \right\rangle =\int_{0}^{1}duC^{B1}\left(m_{b},u\right)
 \left\langle \mathcal{O}_{8g}^{B1}\left(u\right) \right\rangle\,.
\end{equation}
We do not consider QED corrections. At leading order in $\alpha_s$, we find that only the QED current with an $s$ quark contributes, and we arrive at 
\begin{equation}
\label{matchingbLO}
{C_{LO}^{B1}\left(m_{b},u\right)= (-1) \frac{\overline{u}}{u}\, \frac{m_{b}^{2}}{4\pi^{2}}\,\frac{G_{F}}{\sqrt{2}}\lambda_{t}\, C_{8g}} = (-1) \frac{\overline{u}}{u} C_{LO}^{A0}\left(m_{b}\right) \,,
\end{equation}
where we use hardcollinear momentum conservation $\bar n s+\bar n r=m_{b}$ and introduce hardcollinear momentum fraction $u=\frac{\bar n p_s}{m_{b}}$ and $\overline{u}=1-u$.
\begin{figure}
\begin{center}
\includegraphics[width=0.6\textwidth]{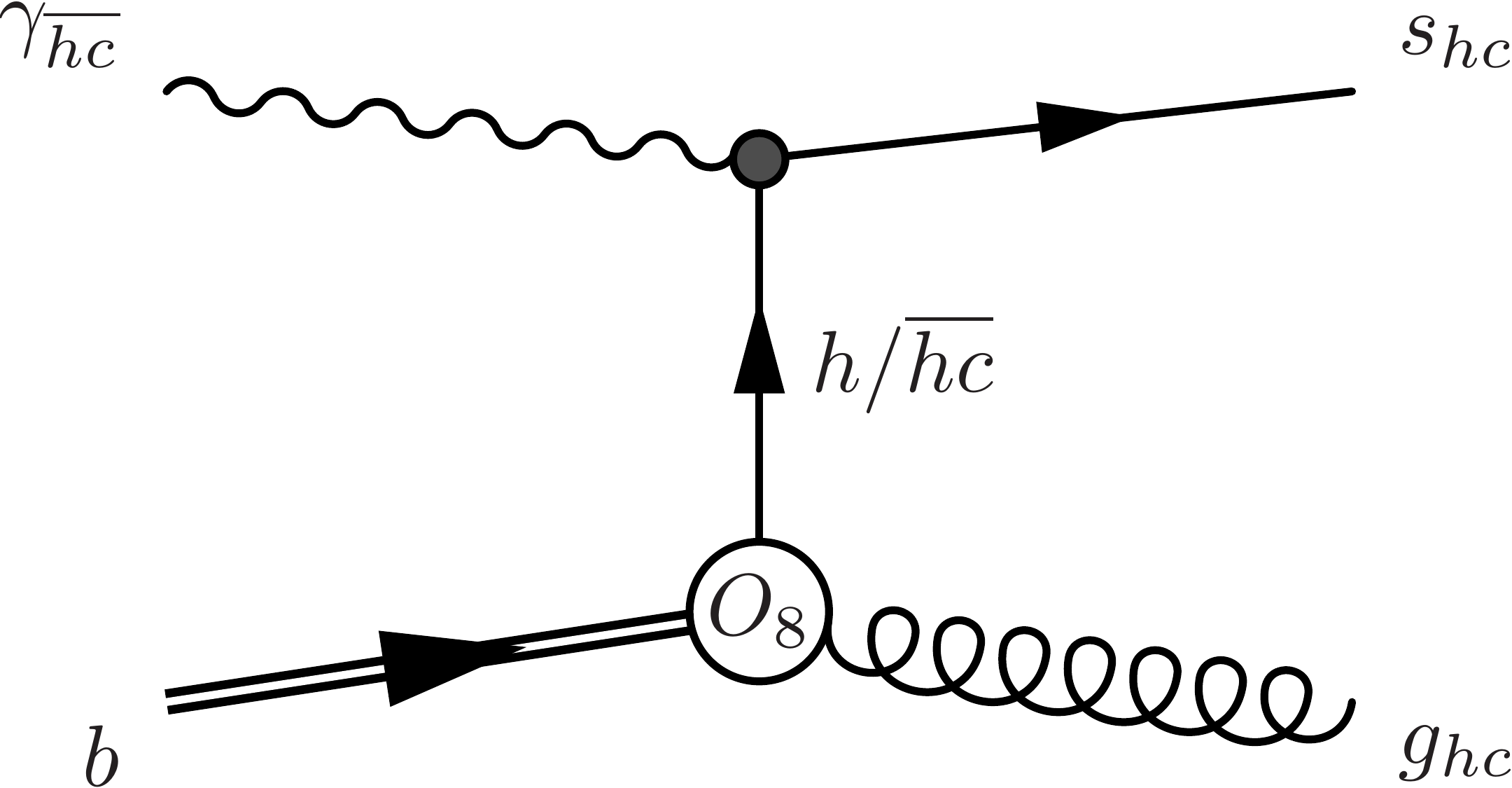}
\caption{\label{fig:FT}The full theory LO diagram which induces an endpoint divergence in SCET, see text.}
\end{center}
\end{figure}

Finally, we compare the different kinematics of the $A$- and $B$-type currents in Figure~\ref{fig:FT}.  The external s-quark carries hardcollinear momentum. Therefore the intermediate propagator is hard. This situation is represented in SCET by the $B$-type current. When the momentum of the external s-quark tends to zero, the propagator becomes anti-hardcollinear and cannot be integrated out -- it must be reproduced by a dynamical field in the low energy EFT. This situation is represented in SCET by the time-order product of subleading Lagrangian and the $A$-type current. The degeneracy in the EFT description is the reason why the SCET develops divergencies in the convolution integrals. 

\vspace{1cm} 

\section{Bare factorisation theorem}

The derivation of the factorisation theorem follows the standard approach. \cite{Vita:2020ckn, Jaskiewicz:2021cfw}. 
We first perform the soft decoupling transformation~\cite{Bauer:2001yt}, but we do not use a new notation for the hardcollinear fields after decoupling. The decay rate is obtained from the imaginary part of the current-current correlator. The states factorise and thus allow taking matrix elements separately in hardcollinear, anti-hardcollinear and soft sectors. 

The common to both $A$- and $B$-type contributions is the anticollinear matrix element of the photon. It is given by a discontinuity of the photon propagator
\begin{align}
- g_{\perp}^{\mu\nu}\,e^2\, J_{\gamma}\left(q^{2}\right) & =\frac{1}{2\pi i }\, {\rm Disc}[ \, i \int d^{4}xe^{iqx}\left\langle 0\right|\, 
T\left[\mathcal{B^{\mu}}_{\overline{hc}\perp}\left(x\right),\mathcal{B^{\nu}}_{\overline{hc}\perp}\left(0\right)  \right] \left|0\right\rangle\,] \\
 & =-g_{\perp}^{\mu\nu} \, e^2\, \delta\left(q^{2}\right)\theta\left(q^0\right) =  -g_{\perp}^{\mu\nu} \, e^2\,\delta^+\left(p^{2}\right) 
\end{align}
Since we are only interested in the photon-final state, the above expression is exact to all orders in perturbation theory. 

\subsection{$B$-type current (direct) contribution}
There are several functions entering the factorisation formula of the direct contribution with the $B$-type current.
The soft function -- the leading power shape function -- is defined as 
\begin{align}
{\cal S}\left(\omega\right) =\frac{1}{2m_{B}}\int\frac{dt}{2\pi}e^{-i\omega t}\left\langle B\right|h\left(t n\right)S_{n}\left(t n\right)S_{n}^{\dagger}\left(0\right)h\left(0\right)\left|B\right\rangle \,,\label{softleadingpower}
\end{align}
or with open indices~\cite{Neubert:1993um,Paz:2009ut}\footnote{We use Greek for spinor indices and Latin for colour indices.}   
\begin{align}
\frac{1}{2m_{B}}\int\frac{dt}{2\pi}e^{-i\omega t}\left\langle B\right| \left[ h_{\alpha}\left(t n\right)S_{n}\left(t n \right) \right]_i  \, \left[ S_{n}^{\dagger}\left(0\right)h_{\beta}\left(0\right)\right]_j  \left|B\right\rangle =\frac{\delta_{ij}}{2\, N_c}\, \left[\frac{1+\slashed v}{2}\right]_{\beta\alpha}{\cal S}\left(\omega\right)\,.
\end{align}

The hardcollinear jet function is a genuine NLP object. In analogy to the LP jet function, we define it as a vacuum matrix element of a product of hardcollinear fields
\begin{align} 
&J\left(p^{2},u,u'\right) = \frac{(-1)}{2 N_c}\, \frac{1}{2\pi} \, \int\frac{dtdt'}{\left(2\pi\right)^{2}}\,\,d^{4}x\,\,e^{-im_{b}\left(ut-u't'\right)+ipx} \label{jetbwithoutopen}\\ 
&\text{Disc}\left[\left\langle 0\right| tr\left[\frac{ 1+\slashed v}{2}(1-\gamma_5) \slashed{\mathcal{A}}_{hc\perp}\left(x\right)  \gamma_{\perp}^{\nu}\chi_{hc}\left(t'\bar{n}+x\right)\overline{\chi}_{hc}\left(t\bar{n}\right)\gamma_{\nu\perp}\slashed{\mathcal{A}}_{hc\perp}\left(0\right) \left(1+\gamma_{5}\right)\right]\left|0\right ] \rangle \right] \,.
\nonumber \end{align}
The field operators are time- or anti-time-ordered according to the Keldysh formalism.\footnote{For a brief summary, see
appendix of Ref.~\cite{Becher:2007ty}.} {The trace is taken both with respect to colour and spinor spaces.}
Using projection properties of the hardcollinear fields and $2\slashed v = \slashed n_+ +\slashed n_-  $  the Dirac algebra in eq.(\ref{jetbwithoutopen}) can be simplified to
\begin{align}
&J\left(p^{2},u,u'\right)=\frac{(-1)}{2 N_c}\,\frac{1}{2\pi}\, \int\,\frac{dtdt'}{\left(2\pi\right)^{2}}\,d^{4}x\,e^{-im_{b}\left(ut-u't'\right)+ipx} \, (d-2)^2\\
&\text{Disc}\left[\left\langle 0\right|tr\left[\frac{\slashed {\bar n}}{4} (1 -\gamma_5) \mathcal{A^{\mu}}_{hc\perp}\left(x\right)\chi_{hc}\left(t'\bar n+x\right)\overline{\chi}_{hc}\left(t\bar n\right)\mathcal{A}_{\mu}^{hc\perp}\left(0\right) \left(1+\gamma_{5}\right)\right]\left|0\right\rangle \right]\nonumber
\end{align}
 where, as mentioned before, the trace is also applied in the colour space.

With these definitions, we find the bare factorisation theorem for the direct contribution
\begin{equation}
\frac{d\Gamma}{dE_{\gamma}}= \mathcal{N}_B\, \int_{0}^{1}du  C^{B1}\left(m_{b},u\right)\int_{0}^{1}du' C^{B1*}\left(m_{b},u'\right)\int_{-p_{+}}^{\overline{\Lambda}}d\omega J\left(M_{B}\left(p_{+}+\omega\right),u,u'\right){\cal S}\left(\omega\right)\label{barefactorB}
\end{equation}
with the prefactor
\begin{equation}
\mathcal{N}_B = [(2 \pi)] \,\, [e^2 Q_s^2]\,\, [\frac{1}{(2\pi)^3\,2\,E_\gamma} 
E_\gamma^2\, 4 \pi] = e^2 Q_s^2 \frac{E_\gamma}{2\pi}
\end{equation}
{The three pieces of the prefactor correspond to the phase space factors of the photon, to its charges and to the redefinition of the jet function with a $2\pi$ factor.}

Finally, we prove to all orders in $\alpha_s$ that the jet-function is symmetric in $u$ and $u'$ up to complex conjugation:
\begin{equation}
J(p^2,u, u') = J^*(p^2, u' , u). \label{symmetryB}
\end{equation}
This can be read off from the factorisation theorem of the direct contribution. The photon energy spectrum is real. The leading power shape function is also real to all orders. This can be shown by complex conjugation of Eq.(\ref{softleadingpower})
and by using translation invariance~\cite{Benzke:2010js}.  Then the jet function inherits the symmetry property given in 
Eq.(\ref{symmetryB}), from the product of the Wilson coefficients, $C^{B1}\left(m_{b},u\right)C^{B1*}\left(m_{b}, u' \right)$, in the convolution integral. An anti-symmetric part of the jet function would cancel out in the convolution integral. 
We emphasise that this property is also valid when the other $B$-type operator with the reversed Dirac structure is present. In particular, the sum of the two mixed terms has this property. In the latter terms, the reduction of the Dirac structure leads to $(4-d)\,(d-2)$, and hence these terms vanish for $d=4$.

The symmetry property is crucial for the refactorisation because it implies that no double subtraction regarding the variables $u$ and $u'$ is needed in the $B$-type (direct) current contribution. This can be seen in the following way. We showed above that the integrand of the convolution integral of the Wilson coefficients and the jet function in the two variables $u$ and $\bar u$ is real, so the complete integrand is symmetric in $u$ and $u'$. Then the subsequent rearrangement is possible (we here only write the convolution variables $u$ and $u'$):
\begin{equation}\label{SymB}
\int_{0}^{1}du C^{B1}\left(u\right)\int_{0}^{1}du' C^{B1*}\left(u' \right) J\left(u,u'\right) = 2\, \int_{0}^{1}duC^{B1}\left(u\right)\int_{u}^{1}du' C^{B1*}\left(u' \right) J\left(u,u'\right) \,.
\end{equation}
 As the endpoint divergence manifests for small $u$ and $u'$, we need to ensure that only the last integral over $u$ is rendered finite by an appropriate subtraction.

At the leading order, the jet function is real, and we find that the jet function is symmetric in $u$ and $u'$. Explicitly, we find using the dimensional  $\overline{\mbox{MS}}$ regulator $(\mu^{2\epsilon}\rightarrow\mu^{2\epsilon}\, \exp(\gamma_E \epsilon) / (4\pi)^\epsilon)$:
\begin{align} 
J\left(p^2,u,u'\right) = C_F \frac{\alpha_s}{4\pi\, m_b}\theta(p^2)\, A(\epsilon)\, \delta(u-u') u^{1-\epsilon} (1-u)^{-\epsilon} \left(\frac{p^2}{\mu^2} \right)^{-\epsilon}\,,
\end{align}
with 
\begin{equation}
A(\epsilon) = (2-2\epsilon)^2\, (1- 1/2\, \epsilon)\,\Gamma(1-\epsilon)^{-1} \exp(\gamma_E \epsilon) = 4 - 10 \epsilon +\mathcal{O}(\epsilon^2)
\label{Aeps}\,.
\end{equation}

We compute the convolution integrals explicitly\footnote{Symmetry of the original integral implies that $\int_u^{1} du' \delta(u-u')= \theta(0)$ with $\theta(0)=1/2$, for $u\in [0,1]$. } 
 using this leading order result for the jet function and also the hard function at leading order, Eq.(\ref{matchingbLO}),
\begin{align}
\frac{d\Gamma}{dE_{\gamma}}|_B &=2\mathcal{N}_B\, \left|C_{LO}^{A0}\left(m_{b}\right)\right|^2  \int_{0}^{1}du \frac{\bar u}{u} 
\int_{u}^{1}du' \frac{\bar u'}{u'}\label{Bfactleading}\\
&C_F A(\epsilon) \frac{\alpha_s}{(4\pi)\, m_b}\, 
\int_{-p_{+}}^{\overline{\Lambda}}d\omega\, {{\cal S}\left(\omega\right)}\left(\frac{m_b (p_+ + \omega)}{\mu^2} \right)^{-\epsilon} u^{1-\epsilon} (1-u)^{-\epsilon} \delta(u-u') \nonumber\\
&=\mathcal{N}_B\, \left|C_{LO}^{A0}\left(m_{b}\right)\right|^{2} \,
C_F \frac{\alpha_s}{(4\pi)\, m_b} \int_{-p_+}^{\overline{\Lambda}} d \omega\, {\cal S}(\omega)\, A(\epsilon)  
  B(3-\epsilon,-\epsilon)\left(\frac{m_b (\omega+p_+)}{\mu^2} \right)^{-\epsilon}\,,\nonumber
\end{align}
where $B(x,y)$ denotes the Beta function. We see that the divergence in the direct contribution is now identified as an endpoint point divergence in the convolution integral of the hard and the jet function in the $u$ integration for $u\ll 1$. 

We emphasise that this endpoint divergence can be regularised within the dimensional regularisation scheme\footnote{We note that we do not confirm the leading order result of the direct contribution of Ref.~\cite{Benzke:2010js} in the dimensional regularisation scheme. In the notation of Ref.~\cite{Benzke:2010js} we get 
\begin{equation}
  F_{88}^{(a)}(E_\gamma,\mu) 
  = \frac{C_F\alpha_s(\mu)}{4\pi}
  \left( \frac{m_b}{2E_\gamma} \right)^2
  \int_{-p_+}^{\bar\Lambda}\!d\omega
  \left( \frac29 \ln\frac{m_b(\omega+p_+)}{\mu^2} + \frac29 \right) {\cal S}(\omega,\mu) \,. \nonumber 
\end{equation}
}.
This leads to additional poles after performing the convolution. Consequently, due to endpoint divergences, the bare factorisation formula is already invalid for the $d\to 4$ limit at the leading order.\\
\newpage

\subsection{$A$-type (resolved) contribution }
For the resolved contribution with the $A$-type current, we start with the time-ordered product 
\begin{align}
&\mathcal{O}_{T\xi q} =i\int d^{d}xT\left[ \mathcal{L}_{\xi q}\left(x\right),\mathcal{O}_{8g}^{A0}\left(0\right)\right] 
\nonumber\\
 &=i\int d^{d}xT\Big[\overline{q}_{s}\left(x_{+}\right) {S_{\overline{n}}(x_+)}  \left(Q_s\, \slashed{\mathcal{B}}_{\overline{hc}\perp} +
\slashed{\mathcal{A}}_{\overline{hc}\perp}\right)
 \left(x\right)\chi_{\overline{hc}}\left(x\right),\nonumber \\
 &\overline{\chi}_{\overline{hc}}\left(0\right) { S_{\overline{n}}^\dagger(0)}   S_n(0)\frac{\slashed n}{2} \slashed{\mathcal{A}}_{hc\perp}(0)\left(1+\gamma_{5}\right) {S_{n}^\dagger(0)} h\left(0\right)\Big] .\label{TproductA}
\end{align}
The operator in the hardcollinear sector contains only gluon fields. Hence the standard leading power gluon jet function appears \begin{align}
-g_{s}^{2}\delta_{ab}g_{\perp}^{\mu\nu}J_{g}\left(p^{2}\right) & = \frac{1}{2\pi i}\, {\rm Disc} \left[ i \int d^{4}xe^{ipx}\left\langle 0\right| T\left[ \mathcal{A}^{a \mu}_{{hc}\perp}\left(x\right),\mathcal{A}^{b \nu}_{{hc}\perp}\left(0\right) \right] \left|0\right\rangle \right].
\end{align} 
 At leading order we find the standard result $J_{g}\left(p^{2}\right) = \delta^+\left(p^{2}\right) $. 
 
Besides photons, there are no energetic particles emitted in the anti-hardcollinear directions. Thus,  the anti-hardcollinear jet function is defined at the amplitude level:
\begin{equation}
\mathcal{O}_{T\xi q}=\int d\omega\int\frac{dt}{2\pi}e^{-it\omega}\left[\overline{q_{s}}\right]_{\alpha}\left(t n\right)\left[\overline{J}\left(\omega\right)\right]_{\alpha\beta}^{a\, \nu\mu}\, Q_s\, \mathcal{B^{\nu}}_{\overline{hc}\perp}\left(0\right)\mathcal{A}_{hc\perp}^{\mu}\left(0\right)\left[h\left(0\right)\right]_{\beta}.
\end{equation}
The anti-hardcollinear jet function can be decomposed as 
\begin{equation}\label{eq:Jbar}
\left[\overline{J}\left(\omega\right)\right]_{\alpha\beta}^{a\, \nu\mu}=\overline{J}\left(\omega\right)\, t^a\, \left[\gamma_{\perp}^{\nu}\gamma_{\perp}^{\mu}\frac{\slashed {\bar n}\slashed n}{4}\right]_{\alpha\beta},
\end{equation}
to all orders.
The other structure $\gamma^\mu_\perp \gamma^\nu_\perp$ does not appear as one can read off from the structure of the $T$ product in eq.~(\ref{TproductA}) and the fact that the gluon and heavy quark fields are only spectators. 
The Dirac structure can then be  simplified at the level of the cross-section with the help of the following relation:
\begin{equation}
\left[\gamma_{\perp}^{\nu}\gamma_{\perp}^{\mu}\frac{\slashed {\bar n} \slashed n}{4}\right]_{\alpha\beta}\left[\gamma_{\perp}^{\mu}\gamma_{\perp}^{\nu}\frac{\slashed n\slashed {\bar n}}{4}\right]_{\alpha'\beta'}
=\left(d-2\right)^{2}\left[\frac{\slashed {\bar n} \slashed n}{4}\right]_{\alpha\beta}\left[\frac{\slashed n \slashed {\bar n}}{4}\right]_{\alpha'\beta'} \,.
\end{equation}  
At leading order, the anti-hardcollinear jet function is given by  
\begin{equation}
\left[\overline{J}\left(\omega\right)\right]_{\alpha\beta}^{a\, \nu\mu}=\frac{t^a}{(\omega +i\,\epsilon) }\left[\gamma_{\perp}^{\nu}\gamma_{\perp}^{\mu}\frac{\slashed {\bar n}\slashed n}{4}\right]_{\alpha\beta}.
\end{equation}

Having defined hardcollinear and anti-hardcollinear functions, we now focus on the soft sector. The operatorial definition of the soft function in position space with open Dirac indices is  
\begin{align}\label{eq:softS}
&{\cal S}_{\alpha\beta,\alpha'\beta'}\left(u,t,t'\right) =\nonumber\\
&= g_{s}^{2}\left\langle B\right|\left[\overline{h}\left(un\right)(1-\gamma_5)\right]_{\alpha'}\left[S_{n}\left(un\right)t^{a}S_{n}^{\dagger}\left(un\right)\right]S_{\bar n}\left(un\right)\left[S_{\bar n}^{\dagger}\left(t'\bar n
+un\right)q_{s}\left(t'\bar n +un\right)\right]_{\beta'} \nonumber\\
 & \times\left[\overline{q}_{s}\left(t\bar n\right) S_{\bar n}^{}\left(t \bar n \right) \right]_{\alpha} S_{\bar n}^{\dagger}\left(0\right)\left[S_{n}\left(0\right)t^{a}S_{n}^{\dagger}\left(0\right)\right]\left[(1 + \gamma_5) h\left(0\right)\right]_{\beta}\left|B\right\rangle\,/\, (2 m_B)\,.
\end{align}
We can now plug in all the objects into the matrix element squared, and we find the resolved contribution
\begin{equation}\label{eq:resolv}
\frac{d\Gamma}{dE_{\gamma}} = \mathcal{N}_A\, \left|C^{A0}\left(m_{b}\right)\right|^{2} { \int_{-p_+}^{\overline{\Lambda}}} d\omega J_{g}\left({ m_{b}}\left(p_{+}+\omega\right)\right)\int d\omega_{1}\int d\omega_{2}\overline{J}\left(\omega_{1}\right)\overline{J}^{*}\left(\omega_{2}\right) {\cal S}\left(\omega,\omega_{1},\omega_{2}\right)\,,
\end{equation}
with the prefactor 
\begin{align}
\mathcal{N}_A = \mathcal{N}_B \equiv \mathcal{N}\,,
\end{align}
and the scalar soft function obtained after contracting spinor indices according to
\begin{align}
{\cal S}\left(u,t,t'\right) & ={(d-2)^2}g_{s}^{2}\left\langle B\right|\overline{h}\left(un\right)\left(1-\gamma_{5}\right)  \left[S_{n}\left(un\right)t^{a}S_{n}^{\dagger}\left(un\right)\right]S_{\bar n}\left(un\right)S_{\bar n}^{\dagger}\left(t' \bar n+u n\right) \\ &\frac{\slashed n \slashed {\bar n}}{4}     q_{s}\left(t' \bar n +un\right) 
 \overline{q}_{s}\left(t \bar n\right)\frac{\slashed {\bar n}\slashed n}{4}S_{\bar n}^{}\left(t \bar n \right)S_{\bar n}^{\dagger}\left(0\right)\left[S_{n}\left(0\right)t^{a}S_{n}^{\dagger}\left(0\right)\right]\left(1+\gamma_{5}\right)h\left(0\right)\left|B\right\rangle\, / \, (2 m_B)\nonumber \,.
\end{align}
The soft function in momentum space, which appears in eq.~(\ref{eq:resolv}), is obtained through the Fourier transform of the position space expression according to \\
\begin{equation}
{\cal S}\left(\omega,\omega_{1},\omega_{2}\right)=\int\frac{du}{2\pi}e^{-iu\omega}\int\frac{dt}{2\pi}e^{-it\omega_{1}}\int\frac{dt'}{2\pi}e^{it'\omega_{2}}{\cal S}\left(u,t,t'\right) \,.
\end{equation}

As the NLP jet function in $u$ and $u'$ variables, the soft function ${\cal S}\left(\omega,\omega_{1},\omega_{2}\right)$ is symmetric in $\omega_1$ and $\omega_2$ up to complex conjugation:
\begin{equation}
{\cal S}\left(\omega,\omega_{1},\omega_{2}\right) = {\cal S}^*\left(\omega,\omega_{2},\omega_{2}\right) \,.
\end{equation}
This property stems from the fact that the gluon jet function is real to all orders. Thus, the soft function inherits 
the symmetry property from the product of the anti-hardcollinear jet functions, $\overline{J}\left(\omega_{1}\right)\overline{J}^{*}\left(\omega_{2}\right)$ in the factorisation formula of the resolved contribution. Any anti-symmetric part would cancel in the convolution integral. This symmetry property implies that within the refactorisation, a double-subtraction regarding the variables $\omega_1$ and $\omega_2$ in the $A$-type current contribution is not needed either. The symmetry implies that the integrand in the convolution integral between anti-hardcollinear jet functions and soft function in the two variables $\omega_1$ and $\omega_2$ is real and symmetric and allows for the following rearrangement of the convolution integral
 \begin{equation}
\int_{-\infty}^{\infty}
 d\omega_{1} \int_{-\infty}^{\infty}
 d\omega_{2}\overline{J}\left(\omega_{1}\right)\overline{J}^{*}\left(\omega_{2}\right) 
{\cal S}\left(\omega_{1},\omega_{2}\right) = 2\, \int_{-\infty}^{\infty} d\omega_{1}\int_{-\infty}^{\omega_1} 
d\omega_{2}\overline{J}\left(\omega_{1}\right)\overline{J}^{*}\left(\omega_{2}\right) 
{\cal S}\left(\omega_{1},\omega_{2}\right), \label{SymA}
\end{equation}
which is motivated by the fact in the resolved contribution. As we will see explicitly in section \ref{RefactLO}, the convolution integral of the jet and shape function is logarithmically divergent for $\omega_{1,2}\to \infty$. 

At leading order, we find the factorisation formula in case of the $A$-type current (resolved) contribution:\footnote{This confirms the leading order result of the resolved contribution of Ref.~\cite{Benzke:2010js} when the asymptotic limit of the soft function is not yet considered.}
\begin{equation}
\frac{d\Gamma}{dE_{\gamma}} = 
2\mathcal{N}\, \left|C_{LO}^{A0}\left(m_{b}\right)\right|^{2}
\int^{\overline{\Lambda}}_{-p_+}
d\omega \delta\left({m_{b}}\left(p_{+}+\omega\right)\right) \int_{-\infty}^{\infty} \hspace{-0.2cm} d\omega_{1}\int_{-\infty}^{\omega_1} \hspace{-0.2cm} d\omega_{2}\, \frac{1}{(\omega_1 - i\epsilon)} 
\frac{1}{(\omega_2 + i\epsilon)}\, {\cal S}\left(\omega,\omega_{1},\omega_{2}\right) \,.\label{factorisationA}
\end{equation}
We keep the soft function unevaluated at this point since this is a nonperturbative object. For $\omega_{1,2} \gg \omega$, the soft function can be shown to be asymptotically constant, which leads to endpoint divergence in the convolution integrals for large $\omega_{1,2}$ (see section \ref{RefactLO}).

\vspace{1cm} 

\section{Refactorisation of the endpoint contribution} 
We here state the three refactorisation relations, which are based on the fact that in the limits $u\sim u' \ll 1$ and 
$\omega_{1}\sim\omega_{2}\gg\omega$ the two terms of the subleading ${\cal O}_8 - {\cal O}_8$ contribution have the same structure. The refactorisation relations are operatorial relations that guarantee the cancellation of endpoint divergences between the two terms to all orders in $\alpha_s$.

\begin{figure}
\begin{center}
\includegraphics[width=0.6\textwidth]{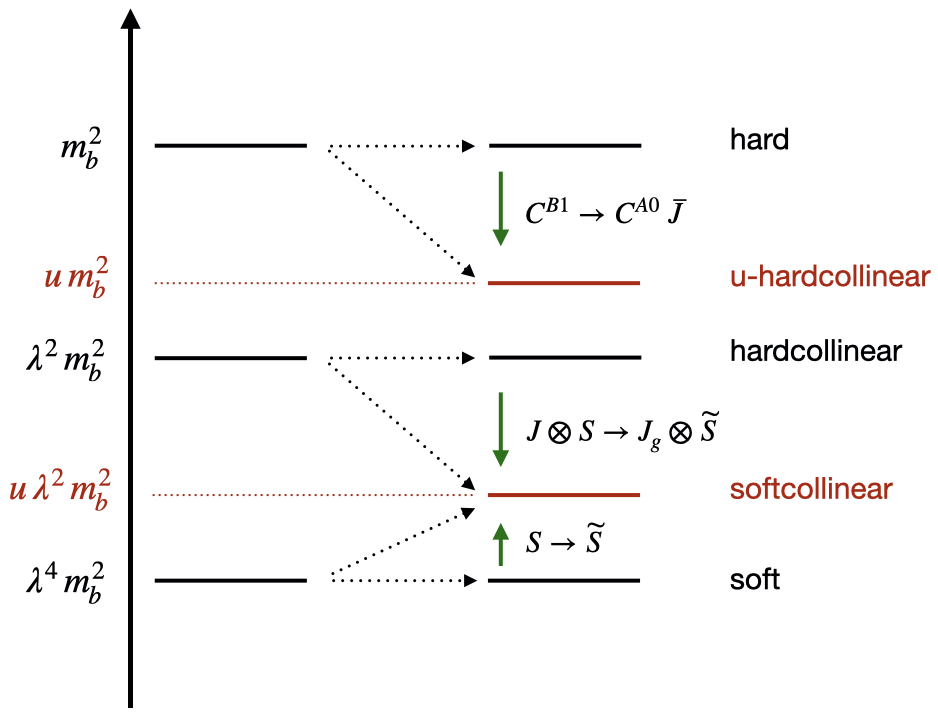}
\caption{\label{fig:scales} Scales relevant to refactorisation of the endpoint divergent contribution. The left part of the diagram represents the standard hierarchy of three scales for SCET$_{\rm I }$. Near the endpoint, when the momentum fraction $u$  is no longer $u\sim O(1)$, i.e. $u\ll1$, we introduce additional, unphysical scales which make it possible to factorise further objects appearing in the bare factorisation theorem. }
\end{center}
\end{figure}

The refactorisation conditions result from the overlap between soft and hardcollinear modes. 
The hierarchy of scales near the endpoint is shown in Fig.~\ref{fig:scales}. We will refer to these overlap modes as 
softcollinear modes. They play a similar role as the $z$-SCET modes introduced in Ref.~\cite{Beneke:2020ibj} to prove the refactorisation of the $B1$-type matching coefficients. The parameter $z$ corresponds to the momentum fraction $u$ in 
the present  analysis. On the one hand, we can think of the softcollinear mode as a limit of hardcollinear mode when the large momentum fraction tends to zero\footnote{{Thus, they do not appear in the leading power problems, where only operators with a single hardcollinear field in each direction occur. }}. On the other hand, the 
softcollinear modes can be understood as a limit of the soft modes when the $n_+k$ momentum component becomes much larger than the remaining components, 
$m_b \gg n_+k \gg \lambda^2 m_b$. We want to emphasise that softcollinear and $u$-hardcollinear modes are not physical but help introduce refactorisation. The softcollinear fields obey the same projection properties and have the same transformation properties regarding gauge invariance as their hardcollinear counterparts. 

\begin{itemize}
\item Following Refs.~\cite{Liu:2019oav,Beneke:2020ibj, Liu:2020wbn}, we find that in the limit $u\to0$, the matching coefficient can be further factorised 
\begin{equation}
{\left\llbracket C^{B1}\left(m_{b},u\right)\right\rrbracket =(-1) C^{A0}\left(m_{b}\right) m_b \overline{J}\left(u m_b \right)\,, \label{eq:RF1}}
\end{equation}
where $\left\llbracket g(u) \right\rrbracket$ only denotes the leading term of a function $g(u)$ in the limit $u \to 0$ 
and without any higher power corrections in $u\ll 1$. The function $ \overline{J}\left(u m_b \right)$, which appears here, is exactly the same radiative jet function (\ref{eq:Jbar}) we introduced before in the context of $A$-type contribution. \\
This refactorisation condition stems from the fact that in the limit $u \to 0  $, the amplitude used in the matching of the $B$-type current can be represented by a time-ordered product  \cite{Beneke:2020ibj},

{\begin{align}\label{eq:refB1}
 C^{B1}\left(m_{b},u\right)
 \left\langle \mathcal{O}_{8g}^{B1}\left(u\right) \right\rangle \Big |_{u\to 0 } = C^{A0}\left(m_b\right) i \int d^{d}x      
 e^{-i \, (nx/2) \, u m_b} \left\langle\;T\left\{ \mathcal{L}_{\xi q_{sc}}^{\left(1\right)}\left(x\right), \mathcal{O}_{8g}^{A0-u}\left(0\right) \right\}\right\rangle \,,
 \end{align}
 of the leading power current $\mathcal{O}_{8g}^{A0-u}(0)=\overline{\chi}_{u-\overline{hc}}(0)\,
{S_{\overline{n}}^\dagger}(0) \,  S_n(0) \,
 \frac{\slashed n}{2}\, \slashed{\mathcal{A}}_{u-hc\perp}(0) \left(1+\gamma_{5}\right)  {S_{{n}}^\dagger}(0) \,     h(0)\,,$ equal to (\ref{eq:A0L}) up to a replacement of the hardcollinear fields by the $u$-hardcollinear fields, and subleading Lagrangian
\begin{equation}
\mathcal{L}_{\xi q_{sc}}^{\left(1\right)}\left(x\right)=\overline{q}_{sc}(x_{+}) 
{S_{{n}}^\dagger}(0) \,{S_{\overline{n}}}(0)\, \left(Q_s\, \slashed{\mathcal{B}}_{u-\overline{hc}\perp} +
\slashed{\mathcal{A}}_{u-\overline{hc}\perp}\right)\chi_{u-\overline{hc}}(x) +h.c.
\end{equation}
The jet-function $ \overline{J}\left(u m_b \right)$ appears after integrating out the $u$-anti-hardcollinear quark fields. We note a close resemblance to the  structure of the resolved contribution, where a similar time-ordered product appears (see Eq. (\ref{TproductA})).}

\item We find the new soft function $\widetilde{{\cal S}}\left(\omega,\omega_{1},\omega_{2}\right)$ which corresponds to the function ${\cal S}\left(\omega,\omega_{1},\omega_{2}\right)$
in the limit $\omega_{1}\sim\omega_{2}\gg\omega$.  
In this limit, we can consider the light soft quarks to be softcollinear $q_{s}\to q_{sc}$. In this function 
$\widetilde{{\cal S}}\left(\omega,\omega_{1},\omega_{2}\right)$ higher power corrections in $\omega/\omega_{1,2} $ are neglected. 

\item In the limit, where the momentum fractions  $u \to 0$ and $u' \to 0$, the jet function \\
$J \left(m_{b}\left(p_{+}+\omega\right),u,u'\right) $ fulfills the following relation
\begin{equation}  \label{refactsquared}
\int_{-p_+}^{\overline{\Lambda}} d\omega
\left\llbracket J \left(m_{b}\left(p_{+}+\omega\right),u,u'\right) {\cal S}(\omega)\right\rrbracket =
 \int_{-p_+}^{\overline{\Lambda}} d\omega
J_{g}(m_b(p_{+}+\omega))
\widetilde{\cal S}(\omega,m_b u ,m_b u')\,,
\end{equation}
where the brackets indicate that the $u \to 0$ {\it and} $u' \to 0$ limits have to be taken and that the hardcollinear quark fields in $J$ are regarded as softcollinear fields, $\chi_{hc} \to q_{sc}$ in accordance with (\ref{eq:refB1}). \\ 
It is crucial that the soft function {$\widetilde{\cal S}\left(\omega,\omega_{1},\omega_{2}\right)$} appears both in the $A$-type contribution in the limit $\omega_{1}\sim\omega_{2}\gg\omega$ and in the $B$-current term if  one expands for small $u$ and $u'$. 

\end{itemize}

Before we proceed, let us comment on the structure of $\widetilde{{\cal S}}$. In the asymptotic regime, where $\omega_{1,2}\gg \omega$, we can match 
the $\widetilde{{\cal S}}$ on the leading power shape function
\begin{align}\label{eq:matchingS}
\widetilde{{\cal S}}\left(\omega,\omega_{1},\omega_{2}\right)  = \int d\omega' K(\omega,\omega',\omega_1,\omega_2)  {\mathcal S}(\omega')\,.
\end{align}
The matching kernel $K(\omega,\omega',\omega_1,\omega_2)$ introduced in (\ref{eq:matchingS}) can be computed perturbatively, i.e. to extract $K$, we can replace $B$-meson state by a $b$-quark in the definition of the soft function and calculate both sides of the equation on the partonic level.   The LP soft function here appears since the limit 
$\omega_{1,2}\to \infty$ is equivalent to the treatment of $t$ and $t'$ as infinitesimal variables in (\ref{eq:softS}) and,  consequently, the soft Wilson lines obtained from decoupling in the anti-hardcollinear direction $S_{\overline{n}}$  cancel. At the same time, the softcollinear quark field produces an additional soft Wilson line associated with the hardcollinear direction $S_{n}$ because we require the softcollinear quark to have the same gauge transformation as a hardcollinear field.
Finally, the structure of the soft  function corresponds to LP shape function ${\mathcal S}(\omega)$. Consistency of the second and third refactorisation conditions, which approach the softcollinear limit from two different directions as shown in Figure~\ref{fig:scales}, leads to
\begin{align}
\int_{-p_+}^{\overline{\Lambda}} d\omega\left\llbracket J \left(m_{b}\left(p_{+}+\omega\right),u,u'\right) {\cal S}(\omega)\right\rrbracket =\int_{-p_+}^{\overline{\Lambda}} d\omega
J_{g}(m_b(p_{+}+\omega))
 \int d\omega' K(\omega,\omega',\omega_1,\omega_2)  {\mathcal S}(\omega').
\end{align}
This relation implies that the kernel $K$ can be obtained from the quark-gluon jet function in the limit when momentum fraction of the quark tends to zero.  Furthermore, it confirms that the kernel $K$ is a perturbative object and that the 
softcollinear scale can be treated perturbatively.

Finally, we note that softcollinear quarks must appear on both sides of the cut. Fermion number conservation implies that only in this case we get a non-vanishing decay rate. Consequently, the endpoint divergences only appear in the limit when both $u$ and $u'$ are small or when $\omega_1$ and $\omega_2$ are large.

\begin{figure}
\begin{center}
\includegraphics[width=0.4\textwidth]{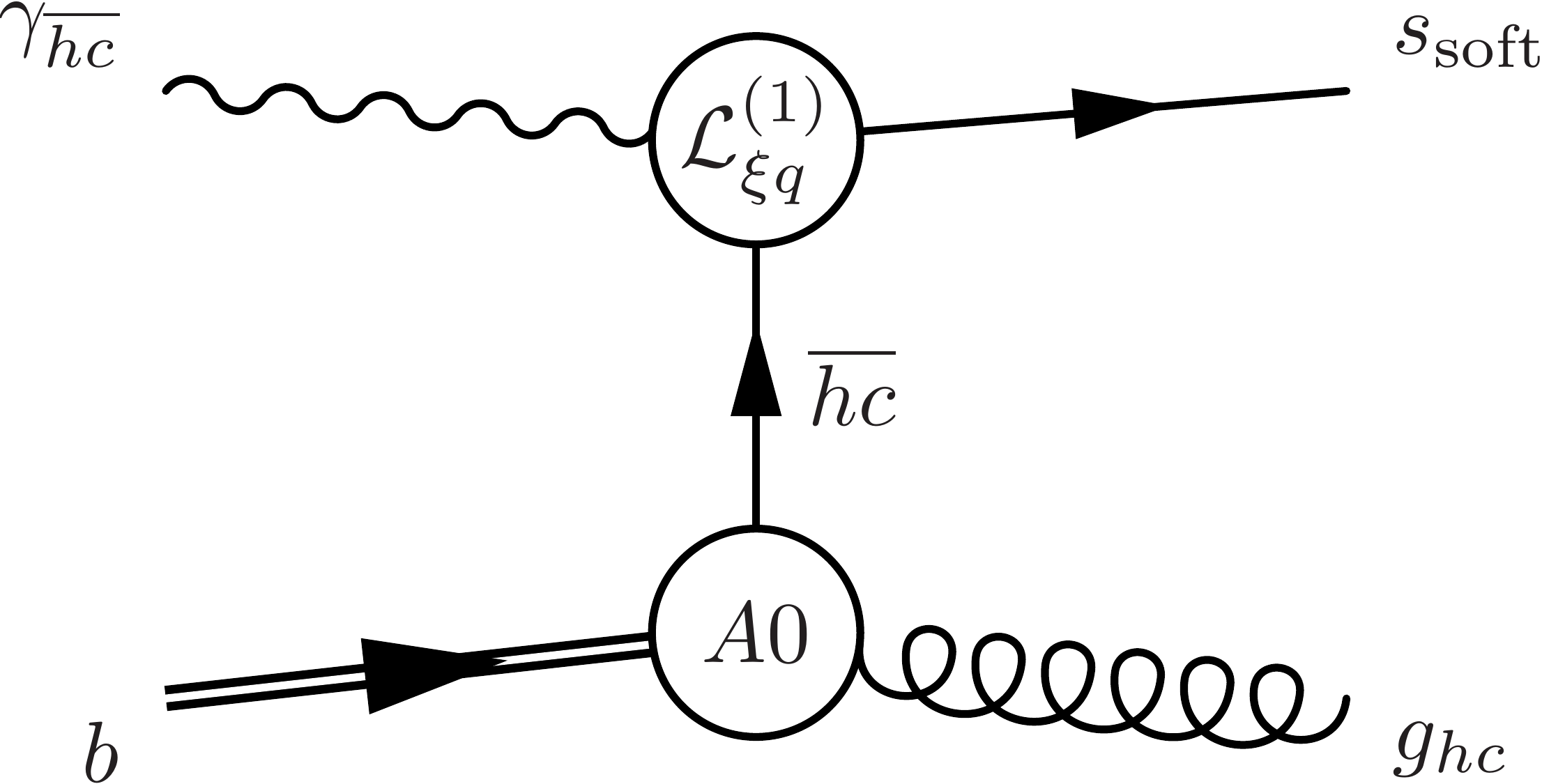}\qquad
\includegraphics[width=0.4\textwidth]{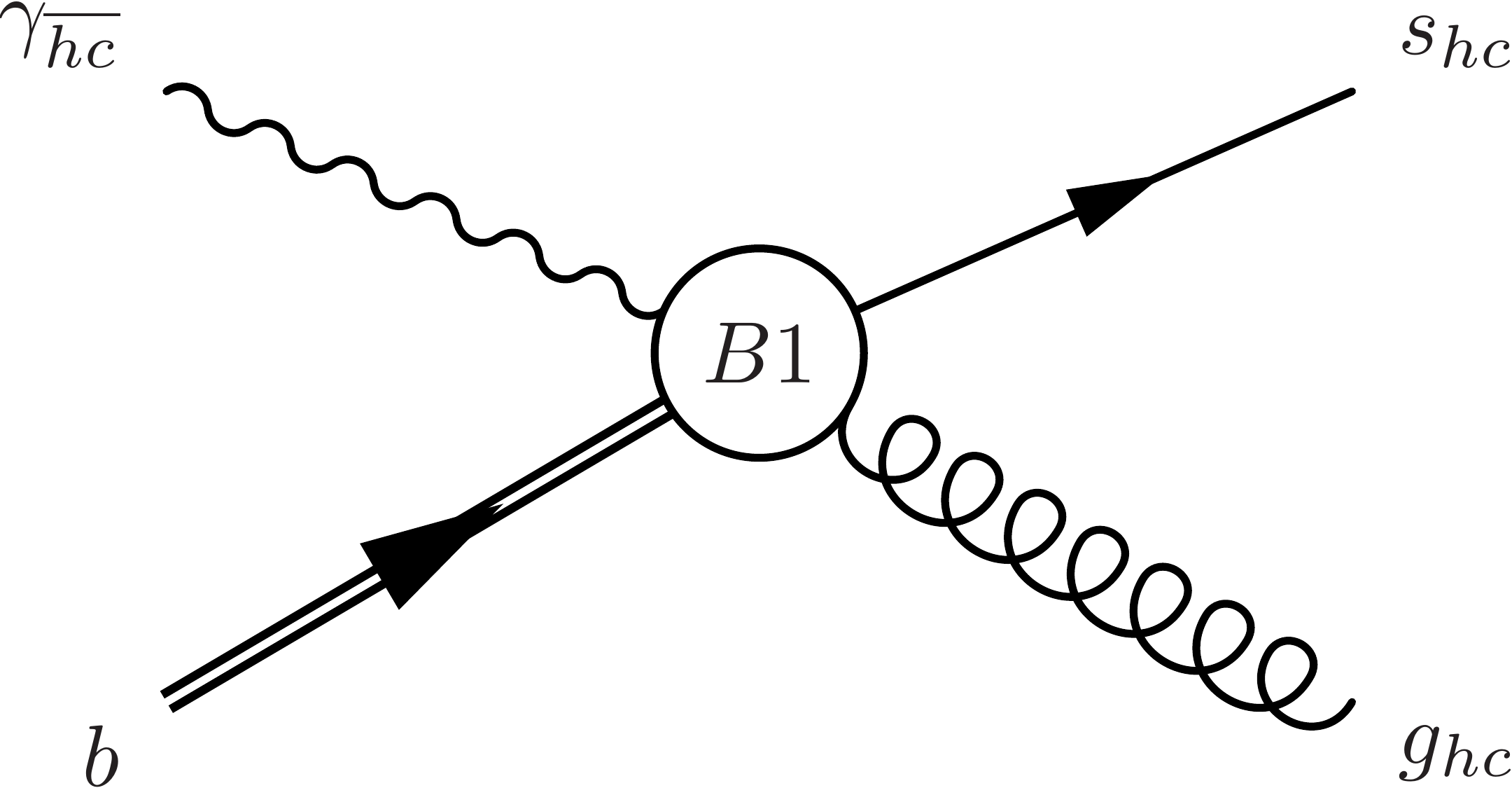}
\caption{\label{fig:SCET} The SCET representations of the full theory diagram in Fig.\ref{fig:FT}, see text.} 
\end{center}
\end{figure}

 Figure~\ref{fig:SCET} shows that the $A$- and $B$-type current have the same structure in the refactorisation limit. On the left, the s-quark is soft and emitted through the insertion of the subleading power Lagrangian. On the right, the s-quark is hardcollinear and emitted directly from the hard $B$-type vertex. When the fraction of the hardcollinear momentum of the s-quark tends to zero, the $B$-type current refactorises into the time-ordered product represented on the left, and both diagrams represent the same full theory configuration. This duality in the description leads to the appearance of the endpoint divergences. A similar problem has already been identified in Refs.~\cite{Beneke:2017vpq,Beneke:2019slt}, in the context of QED corrections in $B_s \to \mu^+ \mu^-$ due to $O_7$ operator at the amplitude level.\\

\subsection{Refactorisation at leading order} 
\label{RefactLO}

Based on the refactorisation conditions, we first discuss the procedure of refactorisation at the leading order. We explicitly verify the conditions using the leading order results. Starting with the last refactorisation condition, we consider the factorisation theorem  of the $A$-type contribution when the soft function is considered in the limit  
$\omega_{1}\sim\omega_{2}\gg\omega$. This asymptotic limit of the soft function can be analysed by means of  semi-perturbative methods~\cite{Bosch:2004th}, where the energetic softcollinear quarks are treated perturbatively, while ordinary soft modes are assumed to be nonperturbative.
In the leading order, this corresponds to the replacement of the softcollinear quarks by a cut propagator. We anticipate the endpoint divergence in the convolution of the soft and the anti-hardcollinear jet functions and use dimensional $\overline{\rm MS}$ regularisation within the calculation. 
We find the following expression of the asymptotic soft function at leading 
order~\cite{Bosch:2004th},:
\begin{equation}
\widetilde{{\cal S}}\left(\omega,\omega_{1},\omega_{2}\right) = C_F A(\epsilon) \frac{\alpha_s}{(4\pi)} \,\, \omega_1^{1-\epsilon} \delta(\omega_1-\omega_2)
\int_\omega^{{\overline\Lambda}} d \omega' \,{{\cal S}(\omega')} \, \left(\frac{(\omega'-\omega)}{\mu^2} \right)^{-\epsilon}\,,
\label{asymptoticsoftfunction}
\end{equation}
which includes the leading power shape function ${\cal S}(\omega)$.  Note that this expression, in principle, receives corrections of higher order in 
$\alpha_s$ and $\Lambda_{\rm QCD}/\omega_{1,2}$, which we do not take into account in the leading order analysis within this section. $A(\epsilon)$ was defined in eq.~(\ref{Aeps})~\footnote{We note  that we do not confirm the leading order result of the asymptotic soft function of Ref.~\cite{Benzke:2010js} in the dimensional regularisation scheme.} 

We convolute the asymptotic soft function with the anti-hardcollinear jet functions for large 
$\omega_1$ and $\omega_2$ only by restricting the limits of the $\omega_1$ integral to $m_b$ and $+\infty$. These integration limits will become clear once we consider the $B$-type current contribution. Starting with the factorisation formula of the $A$ type current given in eq.~(\ref{factorisationA}), the asymptotic contribution of the $A$ type current  reads at leading order:
\begin{align}
\frac{d\Gamma}{dE_{\gamma}}|^{\rm asy}_A &=
2\mathcal{N}\,|C_{LO}^{A0}(m_{b})|^{2}\int_{-p_+}^{\overline{\Lambda}}
\hspace{-0.3cm} d\omega J_g^{LO}(m_{b}(p_{+}+\omega))\int_{m_b}^{\infty}   d\omega_{1} \overline{J}_{LO}(\omega_1)\int_0^{\omega_1} d\omega_{2}\overline{J}^*_{LO}(\omega_2)\,\,\widetilde{{\cal S}}(\omega,\omega_{1},\omega_{2})\nonumber\\
&=\mathcal{N} |C_{LO}^{A0}\left(m_{b}\right)|^{2} \,
 \frac{\alpha_sC_F}{(4\pi)\, m_b} \,  \frac{1}{\epsilon} A(\epsilon)\,  \int_{-p_+}^{\overline{\Lambda}}  \hspace{-0.3cm}    d \omega \,{\cal S}_{LO}(\omega')
 \left(\frac{m_b (\omega+p_+)}{\mu^2} \right)^{-\epsilon}\,.
  \label{Acurrentasymptotic}
\end{align}
The $\frac{1}{\epsilon}$ pole is the manifestation of the endpoint divergence in the resolved contribution in the limit  $\omega_{1}\sim\omega_{2}\gg\omega$. {In the next step, we will see that the specific  choice $m_b$ as a lower limit of the  $\omega_1$  integration  is induced by the refactorisation conditions.} The lower limit in the $\omega_2$ integral can be chosen to be non-negative due to the delta function $\delta(\omega_1 - \omega_2)$.

Now we take the limit $u \to 0$ in the factorisation theorem of the $B$-type current at leading order, which we derived in 
eq.~(\ref{Bfactleading}) before performing the integrals over $u$ and $u'$. This leads to
\begin{align}
\frac{d\Gamma}{dE_{\gamma}} |^{u,u'\to 0}_B 
&=-\mathcal{N}\, \left|C_{LO}^{A0}\left(m_{b}\right)\right|^{2} \,
 \frac{\alpha_sC_F}{(4\pi)\, m_b} \, \frac{1}{\epsilon}\, A(\epsilon) \int_{-p_+}^{\overline{\Lambda}}     d \omega\, {\cal S}_{LO}(\omega) 
\left(\frac{m_b (\omega+p_+)}{\mu^2} \right)^{-\epsilon}. \label{Bcurrentasymptotic}
\end{align} 
This result differs from eq.~(\ref{Acurrentasymptotic}) only by an overall sign. The sum of these two terms is finite and equal to zero. This leading-order result is a special case of the all-order relation, which follows from refactorisation conditions. In the $\omega_{1}\sim\omega_{2}\gg\omega$ (asymptotic) limit of the $A$-type current (with integration limits over $\omega_1$ from $m_b$ to $+\infty$), we exactly single out the same term as in the $u \to 0$ of the $B$-type current up to a minus sign. This reflects the fact that in the limits $u \to 0$ and 
$\omega_{1}\sim\omega_{2}\gg\omega$ the two terms of the subleading ${\cal O}_8 - {\cal O}_8$ contribution have the same structure. Moreover, we   see that with the relations $m_b u = \omega_1$ and $m_b u' = \omega_2$, the $u,u' \to 1$ limit corresponds to the limit $\omega_1,\omega_2 \to m_b$, which fixes the integration limit in the subtraction term of the $A$-type current.

We can summarise the relation we just verified at LO as
\begin{equation} \label{CruxLO}
\frac{d\Gamma}{dE_{\gamma}}|^{\rm asy}_A = (-1) \frac{d\Gamma}{dE_{\gamma}} |^{u,u'\to 0}_B  \,.
\end{equation} 
The refactorisation conditions guarantee that the eq.~(\ref{CruxLO}) holds to all orders in perturbation theory. To make this relation useful for the reshuffling of the factorisation theorem, let us consider an integral of 
$ \widetilde{S}(\omega_1,\omega_2,\omega) \overline{J}(\omega_{1}) \overline{J}^*(\omega_{2})$ over the $\omega_{1,2}\in [0,\infty]$. Since $\widetilde{S}(\omega_1,\omega_2,\omega)$ is expanded for $\omega_{1,2}\gg \omega$, this integral is \emph{scaleless} and equal to zero in dimensional regularisation. We then perform the following manipulations of the integration limits 
\begin{align}
0 &= \int_{0}^{\infty} d\omega_1 \int_{0}^{\infty} d\omega_2 =2 \int_{0}^{\infty} d\omega_1  \int_{0}^{\omega_1} d\omega_2 = 2\left(\int_{0}^{m_b} d\omega_1 +\int_{m_b}^{\infty} d\omega_1\right) \ \int_{0}^{\omega_1} d\omega_2\label{manipulationsoflimits2}.
\end{align} 
In the second step, we made use of the fact that  the integrand is symmetric in $\omega_1$ and $\omega_2$ as we derived to all orders in eq.~(\ref{SymA}). Finally, we split the integration region into two parts suitable for the subtraction. 

The term integrated over $\omega_1$ from $m_b$ to $\infty$ is already in the form suitable for the subtraction of the $A$-type term and equal to (\ref{Acurrentasymptotic}).  To bring the second term into the form of eq.~(\ref{Bcurrentasymptotic}), we perform substitutions  $\omega_1=m_b \,u$ and $\omega_2 = m_b \, u'$, use (\ref{eq:RF1}) to replace  $ \overline{J}(\omega_{1}) $ by the singular part of the $C^{B1}$ matching coefficient and then use the second refactorisation condition to derive 
\begin{align}\label{eq:reltrivial}
&2\mathcal{N} \left|C_{LO}^{A0}(m_{b})\right|^{2} \int_{0}^{m_b} d\omega_{1}\overline{J}_{LO}(\omega_{1})\int_{0}^{\omega_1} d\omega_{2}\overline{J}_{LO}^*(\omega_{2}) 
\int_{-p_{+}}^{\overline{\Lambda}}d\omega J^{LO}_{g}(m_b(p_{+}+\omega))
\widetilde{\cal S}(\omega,\omega_{1},\omega_{2})\nn\\
=\:&2\mathcal{N} \int_{0}^{1}du\left\llbracket C^{B1}_{LO}\left(m_{b},u\right)\right\rrbracket\int_{u}^{1}du' \left\llbracket C^{B1}_{LO}\left(m_{b},u'\right)\right\rrbracket\int_{-p_{+}}^{\overline{\Lambda}}d\omega\left\llbracket J_{ LO}\left(m_{b}\left(p_{+}+\omega\right),u,u'\right){\cal S}_{LO}(\omega)\right\rrbracket   
\end{align}
We rewrite Eq.~(\ref{CruxLO})  using the functions which enter the factorisation theorem:
\begin{align}
&\phantom{-} 2\mathcal{N} \left|C_{LO}^{A0}(m_{b})\right|^{2} \int_{m_b}^{\infty} d\omega_{1}\overline{J}_{LO}(\omega_{1})\int_{0}^{\omega_1} d\omega_{2}\overline{J}_{LO}^*(\omega_{2}) 
\int_{-p_{+}}^{\overline{\Lambda}}d\omega J^{LO}_{g}(m_b(p_{+}+\omega))
\widetilde{\cal S}(\omega,\omega_{1},\omega_{2})  \nonumber\\
=&-2\mathcal{N} \int_{0}^{1}du\left\llbracket C_{LO}^{B1}\left(m_{b},u\right)\right\rrbracket\int_{u}^{1}du' \left\llbracket C_{LO}^{B1}\left(m_{b},u'\right)\right\rrbracket\int_{-p_{+}}^{\overline{\Lambda}}d\omega\left\llbracket J_{ LO}\left(m_{b}\left(p_{+}+\omega\right),u,u'\right){\cal S}_{LO}(\omega)\right\rrbracket .
\label{eq:subtractiontermsLO}
\end{align}
We see with the help of (\ref{eq:reltrivial}) that the fact that  the sum of asymptotic contributions is equal to zero is 
a consequence of our refactorisation conditions. 
It is now clear that these two subtraction terms, which add up to zero, make it possible to reshuffle the factorisation theorem and  cancel the endpoint divergences at the leading order.

\subsection{Bare refactorised factorisation theorem }
The generalisation of the LO order result to all orders is straightforward. Since we are still working in $d$-dimensons with bare objects, we can insert a scaleless expression into the factorisation theorem using the integral manipulations  we performed at LO, see eq.~(\ref{manipulationsoflimits2})

Using the all-orders refactorisation conditions discussed at the beginning of this section, we then can  cast the subtraction term into the following form  with the help of the same manipulations as in the LO case and generalise eq.~(\ref{eq:subtractiontermsLO}) to all orders: 
\begin{align}
0=\:&2\mathcal{N}\left|C^{A0}\left(m_{b}\right)\right|^{2}\int_{-p_{+}}^{\overline{\Lambda}}d\omega J_{g}\left(m_{b}\left(p_{+}+\omega\right)\right)\int_{m_{b}}^{\infty}d\omega_{1}\overline{J}\left(\omega_{1}\right)\, \int_{0}^{\omega_1}d\omega_{2}\overline{J}^*\left(\omega_{2}\right)\widetilde{{\cal S}}\left(\omega,\omega_{1},\omega_{2}\right)   \nn\\
 +\:& 2\mathcal{N}\int_{0}^{1} d u   \left\llbracket C^{B1}\left(m_{b},u'\right)\right\rrbracket    \int_{u}^{1}d u'  \left\llbracket C^{B1*}\left(m_{b},u'\right)\right\rrbracket  \int_{-p_{+}}^{\overline{\Lambda}}d\omega    \left\llbracket J\left(m_{b}\left(p_{+}+\omega\right),u,u'\right) {\cal S}(\omega)\right\rrbracket \:.  \label{Bsubtraction3}
 \end{align}

Starting from the all-order bare factorisation theorem 
\begin{align}
\frac{d\Gamma}{dE_{\gamma}} & = 2\mathcal{N} \left|C^{A0}\left(m_{b}\right)\right|^{2}  \int_{-\infty}^{\infty} d\omega_{1}\overline{J}\left(\omega_{1}\right)\, \int_{-\infty}^{\omega_1} d\omega_{2}\overline{J}^*\left(\omega_{2}\right)
 \int_{-p_{+}}^{\overline{\Lambda}}d\omega J_{g}\left(m_{b}\left(p_{+}+\omega\right)\right)  
{\cal S}\left(\omega,\omega_{1},\omega_{2}\right)\nonumber\\
&+ 2\mathcal{N} \int_{0}^{1}duC^{B1}\left(m_{b},u\right)\,\int_{u}^{1}du'
C^{B1*}\left(m_{b},u'\right)\int_{-p_{+}}^{\overline{\Lambda}}d\omega J\left(m_{b}\left(p_{+}+\omega\right),u,u'\right)
  {\cal S}(\omega)  \end{align}
 and subtracting eq.~(\ref{Bsubtraction3}) we arrive at
\begin{align}\label{eq:bareFT}
\frac{d\Gamma}{dE_{\gamma}}|_{A+B} &= 2 \mathcal{N} \int_{-p_{+}}^{\overline{\Lambda}} 
 \hspace{-0.1cm}   d\omega \bigg\{   J_g(m_{b}(p_{+}+\omega)) \left|C^{A0}\left(m_{b}\right)\right|^{2}   \\
&\times \int_{-\infty}^{\infty}  \hspace{-0.1cm}   d\omega_{1} \,  \int_{-\infty}^{\omega_1}  \hspace{-0.1cm}  d\omega_{2} \overline{J}(\omega_1)\,\overline{J}^*(\omega_2) \left[{\cal S}\left(\omega,\omega_{1},\omega_{2}\right)\nonumber- \theta(\omega_1-m_b)\theta(\omega_2)
\widetilde{\cal S}(\omega,\omega_{1},\omega_{2})\right]\nonumber\\
&+ \int_{0}^{1} \hspace{-0.1cm}   du\int_{u}^{1} \hspace{-0.1cm}  du'  \, \Big[ C_{LO}^{B1}\left(m_{b},u\right)\,
C^{B1*}\left(m_{b},u'\right)\,J\left(m_{b}\left(p_{+}+\omega\right),u,u'\right) {{\cal S}}\left(\omega\right)\nonumber\\
&-   \left\llbracket C^{B1}\left(m_{b},u\right)
\right\rrbracket  
 \left\llbracket C^{B1*}\left(m_{b},u'\right)\right\rrbracket\left\llbracket J\left(m_{b}\left(p_{+}+\omega\right),u,u'\right){\cal S}(\omega)\right\rrbracket  
 \Big]\bigg\}\:,\nonumber
\end{align}
where 
$\left\llbracket J\left(m_{b}\left(p_{+}+\omega\right),u,u'\right) {\cal S}(\omega)\right\rrbracket = J_{g}(m_b(p_{+}+\omega))
\widetilde{\cal S}(\omega,m_b u ,m_b u')$ and $ \left\llbracket C^{B1}\left(m_{b},u'\right)\right\rrbracket\\ = (-1) C^{A0}\left(m_{b}\right)m_b \overline{J}\left(u m_b\right)$.
We note here that the second term effectively restricts the integration range over $\omega_1$ to a finite range in the first line and consequently removes endpoint divergence. Thus these terms need to be added together \emph{before} the $\omega_1$ integral is performed. Similarly, the last term removes the endpoint divergence of the third term, and therefore $u$ integration has to be performed after these two terms are added up.
In addition, we note that the integrals in the first term are finite for large negative values of         
$\omega_1$ and $\omega_2$ due to nonperturbative dynamics~\cite{Benzke:2010js}. At this point, the convolutions integrals in the $A$- and $B$-type contributions are no longer divergent, and we can renormalise the functions entering the factorisation theorem and take the limit $d\to4$.

\subsection{Refactorised factorisation theorem after renormalisation}
We achieved refactorisation at the level of the bare factorisation theorem. It has been pointed out that refactorisation and renormalisation do not commute in general~\cite{Liu:2020tzd,Liu:2020wbn}. Therefore, for the result to be helpful for the resummation of the large logarithms, we must prove that we can express the factorisation theorem in terms of renormalised objects.  To this end, we have to replace bare quantities with renormalised ones. 
The renormalisation of hard matching coefficients is well-established
\begin{align}
C_{\rm bare}^{A0}(m_b) = &\,Z^{A0}(\mu)\, C_{\rm ren}^{A0}(\mu,m_b)\,,\\
C_{\rm bare}^{B1}(u) = &\int_0^1 du'\, Z^{B1}(\mu,u,u')\, C^{B1}_{\rm ren}(\mu.u')\,,
\end{align}
where the one-loop renormalisation factors can be found in Ref.~\cite{Beneke:2005gs}. 
The LP jet function is renormalised according to
\begin{align}
J^{\rm bare}_g(p^2) =  &\int_o^{p^2} d {p'}^2\, Z_{J_g}(\mu,p^2 - {p'}^2)\, J^{\rm ren}_g(\mu,{p'}^2)\,,
\end{align}
with the $Z_{J_g}$ factor given in Refs.~\cite{Becher:2010pd,Banerjee:2018ozf} up to the three-loop order. 
 Similarly, the LP shape function 
 \begin{align}
 S^{\rm bare}(\omega) = &\int d\omega'\, Z_{S}(\mu,\omega-\omega')\, S^{\rm ren}(\mu,\omega')
 \end{align}
 is well-known~\cite{Grozin:1994ni}

Much less is known about NLP objects. The radiative jet function is a notable example which appeared before in the context of $B\to \gamma \ell \nu$~\cite{Bosch:2003fc}.
It has recently been  computed at the two-loop order in Ref.~\cite{Liu:2020ydl}. The most important detail is that  the time-like ($\omega>0$) and space-like $(\omega<0)$ radiative jet functions do not mix under renormalisation 
\begin{align}
\overline{J}^{\;+}_{\rm bare/ren}(\omega) &=\theta(\omega) \overline{J}_{\rm bare/ren}(\omega)\,,\\
\overline{J}^{\;-}_{\rm bare/ren}(\omega) &=\theta(-\omega) \overline{J}_{\rm bare/ren}(\omega)\,,
\end{align}
and 
\begin{align} 
\overline{J}^+_{\rm bare}(\omega) = &\int_0^\infty d \omega' \, Z^+_{\overline{J}}(\mu,\omega,\omega')\,,
\overline{J}^+_{\rm ren}(\mu,\omega')\,,\\ 
\overline{J}^-_{\rm bare}(\omega) = &\int_{-\infty}^0 d \omega' \, Z^-_{\overline{J}}(\mu,\omega,\omega')\,
\overline{J}^-_{\rm ren}(\mu,\omega')\,.
\end{align}
This separation into time-like and spec-like jet functions is necessary since we choose to integrate the subtraction term only over non-negative values of $\omega_{1,2}$. Finally, we define the renormalisation of the NLP soft and jet functions
\begin{align} 
S_{\rm bare}(\omega,\omega_1,\omega_2) = &\int d\omega' d\omega'_1 d\omega'_2\, Z_S(\mu,\omega,\omega',\omega_1,\omega'_1,\omega_2,\omega'_2)\, S_{\rm ren}(\mu,\omega',\omega'_1,\omega'_2)\,,\\
J_{\rm bare}(p^2,u_1,u_2)  =& \int dp'^2 \int_0^1 du'_1\int_0^1 d u'_2\, Z_J(\mu,p^2-p'^2,u_1,u'_1,u_2,u'_2)\, J_{\rm ren}(p'^2,u'_1,u'_2)\,.
\end{align}
These renormalisation kernels are currently unknown. 

We require that $A$- and $B$-type contributions are separately RG invariant (see Ref.~\cite{Liu:2020eqe} for analogous treatment). This leads to the following conditions on the renormalisation kernels
\begin{align}
&| Z^{A0}|^2 \int d\omega \int d\omega_1 \int d\omega_2\, Z_{Jg}(\omega - \omega') 
Z_{\overline{J}}(\omega_1,\omega'_1)\, Z^\dagger_{\overline{J}}(\omega_2,\omega'_2)\,  Z_S(\omega,\omega'',\omega_1,\omega''_1,\omega_2,\omega''_2)\,\nonumber\\ \, = &\delta(\omega'-\omega'')\,\delta(\omega'_1-\omega''_1)\,\delta(\omega'_2-\omega''_2)\,,
 \end{align} 
and 
\begin{align}
&\int_0^1 du_1 \int_0^1 du_2 \int d\omega \, Z^{B1}(u_1,u'_1)\, Z^{B1\dagger}(u_2,u'_2)\, Z_J(\omega-\omega',u_1,u'_1,u_2,u'_2)\, Z_S(\omega-\omega'') \nonumber\\ &= \delta(\omega'-\omega'')\,\delta(u'_1-u''_1)\,\delta(u'_2-u''_2)\,;
\end{align}
and further, RG invariance of the subtraction term leads to 
\begin{align}\label{eq:RGIII}
&| Z^{A0}|^2 \int_0^\infty d\omega_1 \int_0^\infty d\omega_2 \int d\omega \, Z^+_{\overline{J}}(\omega_1,\omega'_1)\, Z^{+\dagger}_{\overline{J}}(\omega_2,\omega'_2)\,Z_{Jg}(\omega - \omega') \, Z_{\widetilde{S}}(\omega-\omega'',\omega_1,\omega''_1,\omega_2,\omega''_2) \nonumber\\ &= \delta(\omega'-\omega'')\,\delta(\omega'_1-\omega''_1)\,\delta(\omega'_2-\omega''_2) \,.
\end{align}
These conditions are sufficient to prove that renormalisation and refactorisation commute and there is no leftover term.

We can now insert the above definitions into eq.~(\ref{eq:bareFT}), 
\begin{align}\label{eq:renFT}
\frac{d\Gamma}{dE_{\gamma}}|_{A+B} &= 2 \mathcal{N} \int_{-p_{+}}^{\overline{\Lambda}} 
 \hspace{-0.1cm}   d\omega \bigg\{   J^{\rm ren}_g (m_{b}(p_{+}+\omega)) \left|C_{\rm ren}^{A0}\left(m_{b}\right)\right|^{2}   \\
&\times \int_{-\infty}^{\infty}  \hspace{-0.1cm}   d\omega_{1} \,  \int_{-\infty}^{\omega_1}  \hspace{-0.1cm}  d\omega_{2} \overline{J}^{+}_{\rm ren}(\omega_1)\,\overline{J}_{\rm ren}^{+*}(\omega_2) \left[{\cal S}_{\rm ren}\left(\omega,\omega_{1},\omega_{2}\right)\nonumber- \theta(\omega_1-m_b)\theta(\omega_2)
\widetilde{\cal S}_{\rm ren}(\omega,\omega_{1},\omega_{2})\right]\nonumber\\
&+ \int_{0}^{1} \hspace{-0.1cm}   du\int_{u}^{1} \hspace{-0.1cm}  du'  \, \Big[ C_{\rm ren}^{B1}\left(m_{b},u\right)\,
C_{\rm ren}^{B1*}\left(m_{b},u'\right)\,J_{\rm ren}\left(m_{b}\left(p_{+}+\omega\right),u,u'\right) {{\cal S}}_{\rm ren}\left(\omega\right)\nonumber\\
&-   \left\llbracket C_{\rm ren}^{B1}\left(m_{b},u\right)
\right\rrbracket  
 \left\llbracket C_{\rm ren}^{B1*}\left(m_{b},u'\right)\right\rrbracket\left\llbracket J_{\rm ren}\left(m_{b}\left(p_{+}+\omega\right),u,u'\right){\cal S}_{\rm ren}(\omega)\right\rrbracket  
 \Big]\bigg\}\,.\nonumber
\end{align}

This is our final result. Endpoint divergences are manifestly absent, assuming one performs the integrals over $\omega_1$ after adding the first and second terms together. Similarly, the integrals over $u$ should be performed after adding the last two lines. This renormalised factorisation theorem allows for a consistent resummation of large logarithms within the
resolved ${\cal O}_8-{\cal O}_8$, using standard RG methods owing to the fact that each object appearing in the above equation is a single scale object. However, a judicious choice of scale might be necessary. 

\newpage

\section{Summary and Outlook} In the present paper, we identified the divergences in the resolved, but also in the direct subleading ${O}_8-{O}_8$ as endpoint divergences which lead to a breakdown of the factorisation theorem already at leading order in four space-time dimensions. The failure of naive factorisation does not allow for consistent separation of scales and, consequently, resummation of large logarithms. 

However, it was recently shown~\cite{Benzke:2020htm} that the resolved contributions still represent the most significant uncertainty in the inclusive $\bar B \to X_s \gamma$ decay. Large scale dependence and also a large charm mass dependence were identified in the lowest order result of the resolved contribution, which calls for a systematic calculation of $\alpha_s$ corrections and RG summation of all resolved contributions~\cite{Benzke:2020htm}. A mandatory input for this task is a well-defined factorisation formula for these subleading corrections. This critical step was established in this paper. The next step consists of computing renormalisation kernels for the NLP soft and jet functions, extracting the anomalous dimensions and solving the RG equations to resum large logarithms. 

Recent intensive studies of the power corrections in collider applications of SCET~\cite{Liu:2019oav,Beneke:2019kgv,Beneke:2020ibj,Moult:2019uhz,Liu:2020tzd,Liu:2020wbn} lead to the development of new techniques that allow for a reshuffling of terms within the factorisation formula so that all endpoint divergences cancel out. We used these new techniques in our flavour application which includes nonperturbative functions typically not present in collider applications of SCET. 
Unlike in the $h \to \gamma \gamma $ decay~\cite{Liu:2019oav}, in the considered SCET$_{\rm I}$ problem, there are no leftover terms present after renormalisation.  

To derive a consistent factorisation theorem, we first established the bare factorisation theorem for the resolved and direct contributions on the operatorial level. Then we derived the all-orders refactorisation conditions applicable to our process. This idea is based on the fact that in certain limits, the two terms of the subleading ${O}_8 - {O}_8$ contribution have the same structure, which guarantees that the endpoint divergences cancel between the two terms to all orders. 
Finally, we proved that we could express the factorisation theorem in terms of renormalised objects so that the result can be used for the resummation of the large
logarithms within the resolved contributions. 


\section*{Acknowledgements} We thank Martin Beneke and Matthias Neubert for their valuable discussions.
RS would also like to thank Mathias Garny and Jian Wang for many discussions on power corrections in SCET and  Mikolaj Misiak for a discussion on the theoretical predictions for $B\to X_s \gamma $. TH is grateful to Michael Benzke for uncounted discussions on the resolved contributions. 
RS is supported by the United States Department of Energy under Grant Contract DESC0012704.
TH is  supported by  the  Cluster  of  Excellence  ``Precision  Physics,  Fundamental
Interactions, and Structure of Matter" (PRISMA$^+$ EXC 2118/1) funded by the German Research Foundation (DFG) within the German Excellence Strategy (Project ID 39083149),  as well as by the BMBF Verbundprojekt 05H2018 - Belle II.  TH also thanks the CERN theory group for its hospitality during his regular visits to CERN where part of the work was done. 

\newpage


\end{document}